\documentclass[prl,twocolumn,showpacs,superscriptaddress]{revtex4}
\usepackage{graphicx}

\begin{document}

\title{Cooperative oscillation of non-degenerate transverse modes in
  an optical system: multimode operation in parametric
  oscillators} 

\author{Axelle Amon}
\affiliation{Universit\'e de Rennes 1, Institut de Physique de Rennes, Campus de Beaulieu, F-35042 
  Rennes, France.}
\affiliation{CNRS, UMR 6251, F-35042 Rennes, France.}
\author{Pierre Suret}
\author{Serge Bielawski}
\author{Dominique Derozier}
\author{Marc Lefranc}

\affiliation{Universit\'e des Sciences et Technologies de Lille,
  Laboratoire de Physique des Lasers, Atomes, Mol\'ecules, CERLA, F-59655
  Villeneuve d'Ascq, France.} \affiliation{CNRS, UMR 8523, FR 2416,
  F-59655 Villeneuve d'Ascq, France.}

\begin{abstract}
  We show experimentally that parametric interaction can induce a
  cooperative oscillation of non simultaneously resonant transverse
  modes in an optical parametric oscillator. More generally, this
  effect is expected to occur in any spatially extended system
  subjected to boundary conditions where nonlinear wave mixing of two
  nonresonant spatial modes can generate a resonant oscillation.
\end{abstract}
\date{\today}
\maketitle 

Sources of coherent radiation such as lasers or optical parametric
oscillators (OPOs) are based on an optical cavity recycling the
generated radiation into the gain medium. The cavity imposes
longitudinal and transverse boundary conditions on the radiation field
and forces it into a superposition of modes, each associated with an
oscillation frequency~\cite{siegman,khanin05:_fundam}. An important
question is whether an optical oscillator can operate simultaneously
on several modes. When mode competition for gain is weakened by
effects such as spatial, spectral or polarization hole-burning, as is
often the case in
lasers~\cite{siegman,khanin05:_fundam,Szwaj98:optical}, intermode
coupling is reduced and multimode operation is common. When instead
modes compete strongly, the lowest-threshold mode prevents the onset
of other modes (as, e.g., in some homogeneously broadened lasers
~\cite{siegman,khanin05:_fundam} and optical parametric
oscillators~\cite{Fabre97,Fabre00:_singlemode}), and single-mode
operation is the rule (``winner-takes-all'' dynamics). This mode
selection problem is in fact not specific to optics and is more
generally relevant to all spatially extended systems where boundary
conditions single out a set of nearly resonant modes which receive energy
from a common source and interact
nonlinearly~\cite{manneville,residori07:_two_farad}.

In the ``winner-takes-all'' case, it is generally believed that
multimode operation can only be observed when several modes have
identical resonance frequencies and
thresholds~\cite{Fabre00:_singlemode,schwob98,vaupel99,ducci01:_patter}.
In optics, such coincidences occur in \emph{degenerate cavities} where
different combinations of longitudinal and transverse mode indices
result in the same oscillation frequency. Except in highly degenerate
cavities such as the plane-parallel, confocal and concentric
two-mirror configurations~\cite{siegman}, mode coincidence is
difficult to achieve and thus multimode operation is generally
considered to be accidental and uncommon.

However, this view assumes a kind of superposition principle, namely
that modes must oscillate individually before they can interact (see,
e.g.,~\cite{manneville}). In this Letter, we show that even in a
``winner-takes-all'' system, the nonlinearity of the gain medium can
induce simultaneous oscillation of transverse modes with well
separated resonance frequencies. More precisely, we experimentally
study multimode operation in a continuous-wave triply resonant optical
parametric oscillator (TROPO), where optical amplification relies on
three-wave mixing~\cite{Fabre97}. Contrary to the common view
that the TROPO is a single-mode
device~\cite{Fabre97,Fabre00:_singlemode}, except in degenerate
configurations~\cite{schwob98,vaupel99,ducci01:_patter}, we provide
experimental evidence of mode 
coexistence between two monomode resonances. Moreover, we observe
multimode operation in many more geometries than predicted from mode
coincidences, in particular far from transverse
degeneracy. Our findings are consistent with predictions of
Ref.~\cite{amon03:_burst} and are similar to recent observations in
Faraday experiments in hydrodynamics~\cite{residori07:_two_farad}.
This suggests that the effect we describe here in an optical system
may more generally occur in systems which combine a strong nonlinear
interaction with tight boundary conditions.

In spite of its simplicity, the TROPO has a rich dynamics, still only
partially understood. It has been predicted theoretically that the
monomode TROPO displays a Hopf bifurcation and a period-doubling
cascade leading to chaos~\cite{lugiato88}. Experimentally, fast
periodic instabilities have been observed in TROPOs at frequencies in
the range 1--300 MHz~\cite{richy95,suret01b}. They occur alone or
combined with slow thermo-optic instabilities~\cite{suret00,suret01a},
giving then rise to bursting oscillations~\cite{amon03:_burst}, and
can follow a transition to chaos~\cite{amon04:_chaos}. However, fast
oscillations can be observed at weaker pumps, smaller signal frequency
detunings and with smaller frequencies than predicted for the monomode
instability~\cite{amon-2007,suret01b,amon03:_burst}.

Using a degenerate two-mode TROPO model~\cite{schwob98}, it was
conjectured in~\cite{suret01b} that these oscillations result from the
interaction of transverse modes, and shown in~\cite{amon03:_burst}
that two transverse modes with disjoint resonance curves can interact
when their resonance frequencies approximately add up to the pump
frequency, generating oscillations at frequencies much higher than
cavity bandwith. Conversion of a pump photon into two photons in the
same mode is then strongly repressed, leaving the cross-conversion
channel (one photon in each of the two modes) as the dominant process.
Roughly speaking, the OPO views the two combined signal-idler modes as
a pair of distinct signal and idler fields, and can thus utilize the
fact that parametric interaction only constraints the sum of the
generated frequencies, not their individual values.

\begin{figure}[htbp]
  \centering
  \includegraphics[width=2.8in]{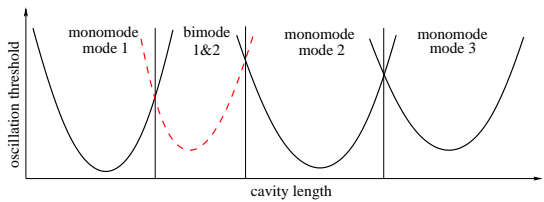}
  \caption{Schematic representation of the resonance spectrum showing
    oscillation threshold of various regimes as a function of cavity length.
    Between two single-mode resonances, there exist two-mode
    resonances where the two modes of the neighboring resonances
    oscillate cooperatively.}
  \label{fig:resonance}
\end{figure}

An important result of Ref.~\cite{amon03:_burst} is that an exact
expression for the two-mode threshold can be computed, and that it is
identical to that of a virtual single-mode regime with an effective
coupling constant that depends on mode overlap. This allows one to
study the mode selection process in a unified way by considering
simultaneously single-mode and two-mode resonances, in which the two
neighboring modes oscillate cooperatively (Fig.~\ref{fig:resonance}).
Because multimode operation is thus allowed to occur without transverse
mode degeneracy, it is expected that such regimes can be observed in a
large number of cavity configurations.

\begin{figure}[htbp]
  \centering
  \includegraphics[width=3.2in]{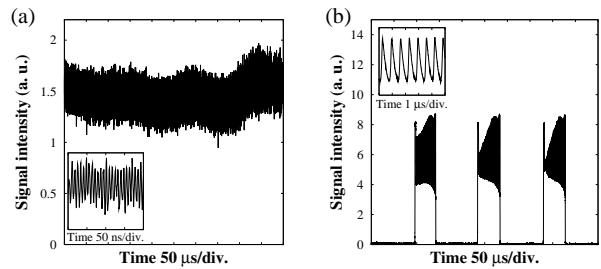}
  \caption{Signal intensity vs. time in two configurations. (a) $L=47$
    mm; threshold 10 mW, $f=130$ MHz~; (b) $L=77$ mm; threshold 33 mW,
    $f=1.2$ MHz--$2.7$ MHz.}
  \label{fig:signals}
\end{figure}

The experimental setup is similar to that used in previous
experiments~\cite{suret01b,suret00,suret01a,amon03:_burst,amon04:_chaos}.
A 15-mm-long KTP crystal cut for type-II phase matching is enclosed
inside a cavity of length $L$ varying between 30 and 100 mm, delimited
by two mirrors with a radius of curvature of 50 mm. Cavity length is
not actively stabilized. Cavity finesses at 532 nm, the wavelength of
the frequency-doubled Nd:YAG pump laser, and at 1064 nm, near which
the signal and idler fields are generated, are 50 and 550,
respectively. The regimes discussed here have been obtained at a pump
power of 3.5 W, with a parametric threshold ranging from 10 mW to 35
mW depending on cavity length. The output pump beam had a TEM$_{00}$
structure. In most configurations, scanning the cavity through one
free spectral range with a PZT allowed us to observe fast oscillations
in the output intensity of the OPO at some cavity lengths, at a
frequency $f$ varying typically from 1 to 300 MHz, combined or not
with opto-thermal oscillations (Fig.~\ref{fig:signals}).

\begin{figure}[htbp]
  \centering
  \includegraphics[width=2.8in]{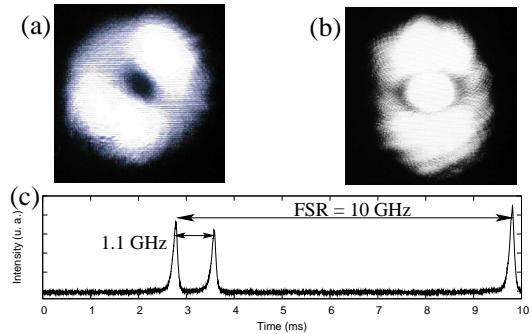}
  \caption{Transverse intensity profile of the signal field recorded
    with a standard CCD camera operating at 50 frames/s. (a) $L=46.3$
    mm, $f=15$ MHz; (b) $L=37$ mm, $f=31.5$ MHz. (c) Intensity of the
    OPO signal at the output of a PF analyser which cavity length is
    swept with time. The transmission presents two maxima over one FSR
    of the PF cavity (10 GHz). The two modes are separated by at least
    1.1 GHz (for OPO cavity: $L=47$ mm, FSR: 2.5 GHz).}
  \label{fig:camera}
\end{figure}

During fast oscillations, the transverse profiles of output signal
beam observed with a CCD camera suggest the
presence of several transverse modes (Fig.~\ref{fig:camera}(a) and
(b)). However, the pattern displayed by the camera is averaged over 20
ms, which does not allow us to distinguish between true mode
coexistence and fast periodic alternation~\cite{suret01a}.
Fig.~\ref{fig:camera}(c) shows a mode analysis using a scanning
P\'erot-Fabry interferometer (PF) of the output signal in a putative
multimode regime. It displays two peaks separated by approximately 1.1
GHz, which is much larger than the cavity bandwith. No mode hopping
can be seen on the output signal during the scan so that this again
suggests coexistence of well-separated modes, but is not conclusive
because the two modes are resolved at different times of the scan.

\begin{figure}[htbp]
  \centering
  \includegraphics[width=3.4in]{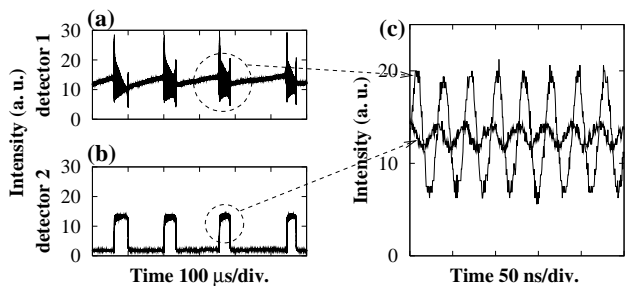}
  \caption{Time traces recorded by two detectors located at different
    points of transverse profile. (a) Detector 1; (b) detector 2;
    (c) superposition of the two out-of-phase time traces during a
    burst of fast oscillations. $L=47$ mm.}
  \label{fig:twodect}
\end{figure}

To obtain convincing evidence of mode coexistence, we have therefore
monitored the output intensity of the signal field in two different
locations of the transverse plane using fast detectors, although this
method probably prevents us to detect very large beat frequencies. A
monomode transverse profile is described by a single time-dependent
amplitude: the time traces of the two detectors should then be
proportional to each other. Fig.~\ref{fig:twodect} shows the two time
traces in a regime where bursts of fast oscillations are observed. The
two traces have very different waveforms and are out of phase during
the fast oscillations, which shows the presence of at least two modes.
This clearly confirms that fast oscillations originate in the
interaction of two or more transverse modes.

\begin{figure}[htbp]
  \centering
  \includegraphics[width=3.2in]{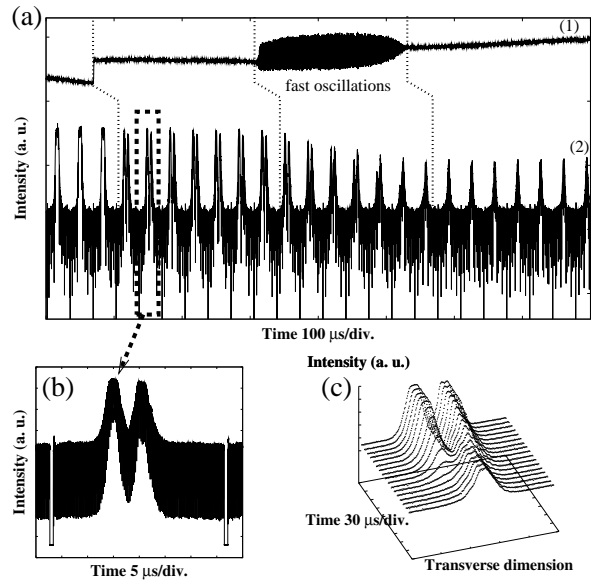}
  \caption{(a) Traces 1 and 2 show total signal intensity as a function of
    time and the raw video signal from the CCD array,
    including interpixel and interframe pulses. Trace 2 is lagging
    behind trace 1 by the CCD transfer time (30 $\mu$s), as indicated
    by dashed lines. (b) The video signal between two interframe
    pulses displays a one-dimensional section of the beam profile
    acquired at the beginning of the segment, superimposed with
    interpixel pulses. (c) Succession of 1D profiles in the
    oscillation zone, extracted from trace 2 by suitable filtering,
    showing gradual transition from a two-hump to a single-hump
    profile.}
  \label{fig:film}
\end{figure}

To study how multimode oscillations are switched on and off when
cavity length is varied, we have monitored the evolution of the beam
spatial profile as the uncontrolled cavity was slowly drifting due to
thermal effects. Sufficient spatial and temporal resolutions could be
simultaneously achieved by using a linear CCD array of 256 pixels, with
an acquisition rate of 33,000 profiles/second. Fig.~\ref{fig:film}
shows a typical observation made in the same configuration as in
Fig~\ref{fig:twodect}. This allows us to correlate qualitative changes
in the total output intensity with modifications in the beam
transverse profile.

In Fig.~\ref{fig:film}, the following events can be seen from left to
right. First there is a mode hop, with a sudden jump in the total
intensity and a transition from a one-hump transverse profile to a
two-hump one. After the mode hop, the output intensity and the
two-hump profile remain approximately constant during 200 $\mu$s.
Then, the system switches abruptly to finite amplitude oscillations,
with no notable change in the transverse profile. As cavity drifts
through the oscillation zone, the transverse profile changes gradually
from a two-hump profile to a one-hump one until oscillations disappear
through an inverse supercritical Hopf bifurcation. Finally, intensity
and the one-hump profile remain constant until the end. This scenario
indicates that fast oscillations occur between two cavity lengths
where two different transverse modes are dominant, as predicted
theoretically~\cite{amon03:_burst}, and that composite profiles are
observed in the unstable zone. A singular value decomposition analysis
of these profiles confirms that they are linear superpositions of the
two monomode profiles.

According to
Ref.~\cite{amon03:_burst}, two-mode regimes should be much
easier to observe than if transverse degeneracy were required for mode
interaction. In order to estimate the prevalence of fast multimode
oscillations in our experimental system, including far from
confocal or concentric geometries, we have thus conducted a systematic
search near cavity lengths $L = 46$, 63, 71, 77 mm. Multimode behavior
has been easily observed in each case.

\begin{figure}[htpb]
  \centering
  \includegraphics[width=3.2in]{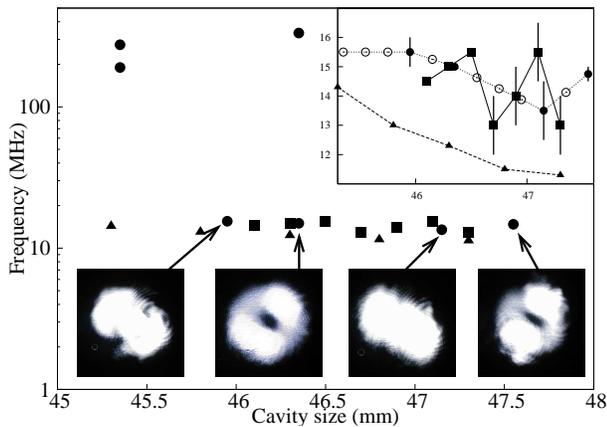}
  \caption{Frequencies of multimode oscillations observed in three
    scans (each associated with a different symbol) of cavity length
    in the 45--48 mm range. Inset shows data points with frequencies
    between 10 and 15 MHz, with a line passing through data points
    of a given scan. Filled (open) symbols denote configurations where
    oscillations were observed (not observed). When different
    oscillation regimes were observed in the same configuration, the
    frequency range is indicated by a vertical bar. Transverse
    profiles observed in the scan associated with circle symbols are
    also shown. }
  \label{fig:scan}
\end{figure}

For example, Fig.~\ref{fig:scan} displays the result of several
cavity-length scans carried out near the configuration of
Figs.~\ref{fig:camera}-\ref{fig:film}, by increments of 200 $\mu$m or
500 $\mu$m. For each configuration, cavity detuning was swept over one
free spectral range to find whether or not fast oscillations could be
observed for some detuning. The data of Fig.~\ref{fig:scan} clearly
show that multimode behavior can be observed in almost all
configurations studied. Surprisingly, transverse profiles almost
always resemble that of low-order modes. Oscillation frequencies
gather around 10--15 MHz, with a few points in the 100--300 MHz range.
For cavity lengths $L=63$, $71$ and $77$ mm, most probable frequencies
were in the 6--15 MHz, the 3--8 MHz and the 1--3 MHz ranges,
respectively, but frequencies well outside these ranges have also been
observed. Thus, lower frequencies are seemingly favored as cavity
length is increased but we do not yet have an explanation for this
behavior.

Note that the degenerate model whose
predictions~\cite{amon03:_burst,suret01b} motivated our experiments is
not suitable to describe a type-II OPO, where signal and idler fields
have orthogonal polarizations. The study of the non-degenerate
multimode model is difficult and is currently in progress. As in the
degenerate case, cross-conversion channels involve signal and idler
fields with different transverse structures, but which now become
resonant in symmetrical pairs because of the small difference between
the periods of the signal and idler frequency combs.

In conclusion, we have shown that the fast oscillations observed in
TROPOs are generated by the coexistence of interacting transverse
modes. The surprising prediction made in Ref.~\cite{amon03:_burst}
that multimode oscillations do not require mode degeneracy also holds
in our system, as confirmed by systematic scans of the cavity lengths
which revealed a high probability of occurence. By monitoring the
transverse structure of the beam as cavity length was slowly swept, we
have also evidenced the occurence of fast oscillations due to mode
coexistence between two resonances corresponding to two different
transverse modes. These findings are important because they show that
two-mode regimes can have a lower threshold than invidual modes and
thus must be taken into account to determine which operating mode is
selected in OPOs~\cite{eckardt91,debuisschert93}.

Finally, this phenomenon should not be specific to OPOs or even
optical systems. Similar observations have been reported in
hydrodynamical Faraday experiments~\cite{residori07:_two_farad}, where
the nonlinearity is formally equivalent to the parametric interaction
in an OPO. In fact, the simplicity of the basic ingredient, namely
that the nonlinear interaction of two non-resonant modes generates a
resonant oscillation that sustains multimode behavior, suggests that
cooperative multimode oscillations may occur in any system where a
nonlinear wave mixing interacts with strong boundary conditions.

We are most grateful to Jaouad Zemmouri for his contribution in the
initiation of this research project and for stimulating
discussions. CERLA is supported by the Minist\`ere charg\'e de la
Recherche, R\'egion Nord-Pas de Calais and FEDER.

\end{document}